# Design and optimization of optical modulators based on graphene-on-silicon nitride microring resonators

Zeru Wu, Yujie Chen*, Tianyou Zhang, Zengkai Shao, Yuanhui Wen, Pengfei Xu, Yanfeng Zhang, and Siyuan Yu

State Key Laboratory of Optoelectronic Materials and Technologies, School of Electronics and Information Technology, Sun Yat-sen University, Guangzhou 510275, China
*E-mail: chenyj69@mail.sysu.edu.cn (Yujie Chen)

**Abstract**
In order to overcome the challenge of obtaining high modulation depth due to weak graphene-light interaction, a graphene-on-silicon nitride ($SiN_x$) microring resonator based on graphene's gate-tunable optical conductivity is proposed and studied. Geometrical parameters of graphene-on-$SiN_x$ waveguide are systematically analyzed and optimized, yielding a loss tunability of 0.04 dB/μm and an effective index variation of 0.0022. We explicitly study the interaction between graphene and a 40-μm-radius microring resonator, where electro-absorptive and electro-refractive modulation are both taken into account. By choosing appropriate graphene coverage and coupling coefficient, a high modulation depth of over 40 dB with large fabrication tolerance is obtained.

## 1. Introduction

Graphene photonics has received a lot of interest due to its outstanding electrical and optical properties of ultrahigh carrier mobility [1], wide operational bandwidth [2], and electrically tunable conductivity [3]. graphene-based photonics devices such as high-speed photodetectors [4,5], electro-optical modulators [6,7], optical polarizer [8] have been demonstrated.

Integrating graphene with passive optical waveguide structures is an effective way to enhance the interaction between light and the graphene sheet [9]. However, the overlap between the evanescent field and graphene is not strong enough to obtain a significant change of optical absorption [10]. To extend the graphene-light interaction, exploiting resonant-enhanced cavity structure is a promising approach and several silicon cavity integrated graphene-based photonics devices with enhanced performance have been reported [11–15]. Compared to silicon waveguides, silicon nitride ($SiN_x$) waveguides have much lower losses, broader transparency windows, smaller thermal-optic effects, and larger fabrication tolerance [16]. Therefore, graphene-based device integrated on $SiN_x$ platform may afford better performance [17–20].

In this work, a comprehensive study on the design optimization of graphene-SiN$_x$ microring modulator is presented. This involves firstly a detailed analysis of optimum waveguide parameters for enhanced grapheme-light interaction. Based on the optimized waveguide structure, several design principles for graphene-based SiN$_x$ microring modulator devices are proposed. These design rules are then applied to a 40-μm-radius microring modulator to produced optimized device designs that can provide a high modulation depth of over 40 dB.

## 2. Simulation of graphene-SiN$_x$ waveguides
### 2.1 Modelling methodology for simulation

Our simulations are performed using the finite differential eigonmode solver (FDE) and graphene is characterized using a surface conductivity model. The surface conductivity of a single layer of graphene can be derived from the Kubo formula [21],

$$\sigma(\omega,\mu,\Gamma,T) = \frac{ie^2(\omega+i2\Gamma)}{\pi\hbar^2}\left[\int_0^\infty \xi\left(\frac{\partial f_d(\xi)}{\partial \xi}-\frac{\partial f_d(-\xi)}{\partial \xi}\right)d\xi - \int_0^\infty \frac{f_d(-\xi)-f_d(\xi)}{(\omega+i2\Gamma)^2-4(\xi/\hbar)^2}d\xi\right] \quad (1)$$

where $\omega$ is the radian frequency of incident light, $\hbar$ the reduced Plank's constant, $\mu$ the chemical potential that can be electro-statically controlled, $\Gamma$ the scattering rate, $T$ the temperature, and $e$ is the charge of an electron. $f_d(\xi) = (e^{\frac{\xi-\mu}{k_B T}}+1)^{-1}$ is the Fermi-Dirac distribution, in which $k_B$ is the Boltzmann's constant. The first and second integral term represent intraband and interband conductivity, respectively. Here $\sigma_0 = e^2/4\hbar = 60.85\,\mu S$ is the minimum conductivity in the optical-frequency regime.

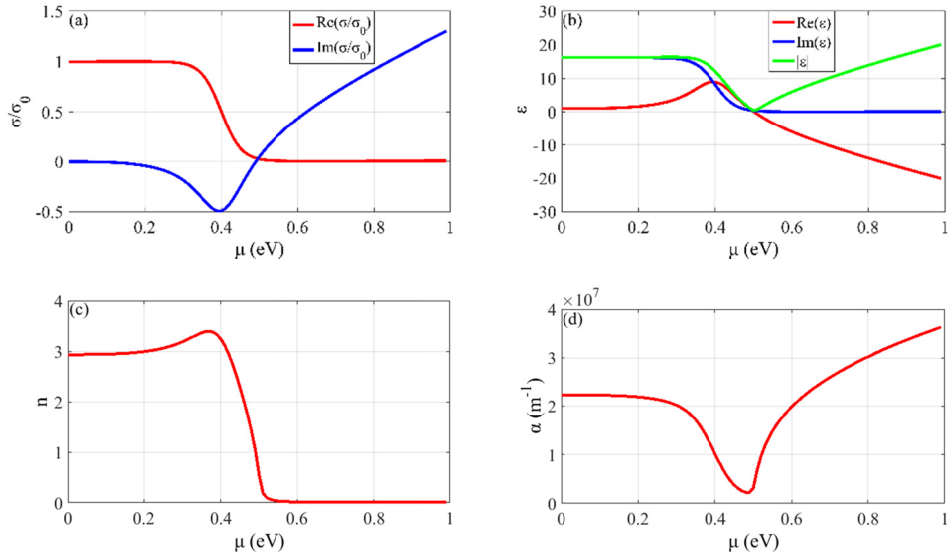

Fig. 1. Electrical and optical properties including (a) conductivity, (b) permittivity constant, (c) refractive index, and (d) absorption coefficient of single layer graphene as a function of chemical potential for $\lambda = 1.55\,\mu m$, $\Gamma = 0.08\,eV$ and $T = 300\,K$.

The following calculations are based on the incident light with $\lambda = 1.55$ μm, $\Gamma = 0.08$ eV, and $T = 300$ K. Optical and electrical properties including conductivity, permittivity constant, refractive index, and absorption coefficient as a function of chemical potential of graphene are shown in Fig. 1. The optical and electrical properties can be dramatically changed via chemical potential tuning, which can be easily controlled by an applied voltage. Note that when chemical potential increases, the sign of imaginary part of permittivity constant can shift from positive to negative, implying that graphene's properties can be modified from dielectric-like to metallic-like [22].

**2.2 Analysis, design, and optimization of graphene-SiN$_x$ waveguides**

Effect of waveguide dimensions and polarization of guided modes on light propagation in a SiN$_x$ waveguide is investigated firstly. The refractive indices of silicon nitride and a 3-μm-thick silicon dioxide (SiO$_2$) substrate layer are 1.95 and 1.46, respectively [16]. The electric field distribution of a quasi-TE guided mode is plotted in Fig. 2(a). To improve electroabsorption modulation efficiency, the optical power confined at the interface between the graphene sheet and the waveguide should be maximized [23]. We perform an exhaustive parameter sweep of geometrical parameters of an air-clad SiN$_x$ waveguide. After several algebraic steps, the single mode condition and corresponding optical power confinement (defined as the ratio of power in the vicinity of the top surface to the total power in the waveguide) is illustratively shown in Figs. 2(b) and 2(c). Power confinement of the evanescent field in a quasi-TE mode waveguide is much larger than that in a quasi-TM mode waveguide. For a single TE mode waveguide with a cross-section of 1.09 μm × 0.33 μm, power confinement reaches to 3.6% while the maximum power confinement for a single TM mode waveguide is 0.80%, with a cross-section of 0.61 μm × 0.54 μm. Thus, to improve electroabsorption modulation efficiency, a quasi-TE mode waveguide with a larger overlap between graphene layer and propagating optical mode is preferable.

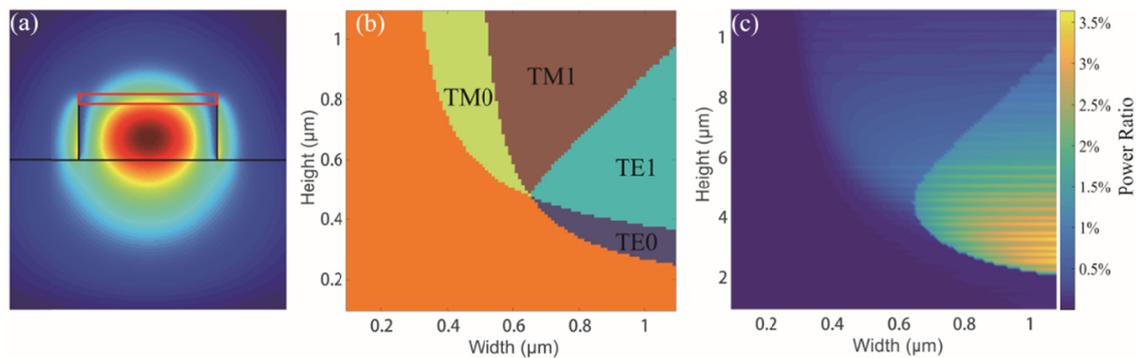

Fig. 2 (a) Cross-section of the SiN$_x$ waveguide with an overlay of electric field intensity, (b) single mode condition of the SiN$_x$ waveguide, and (c) power confinement of the evanescent field in the vicinity of waveguide surface versus geometrical parameters.

After optimization, a TE mode waveguide with a cross-section of 1.09 μm × 0.33 μm is selected for further simulation. Effective index and propagation loss of the waveguide as a function of chemical potential of graphene is shown in Fig. 3(a). It is very important to be aware that both effective index and loss have steep changes around 0.4 eV ~ 0.6 eV. A loss tunability $\Delta\alpha$ of about 0.04 dB/μm and an effective index change $\Delta n_{eff}$ of 0.0022 can be achieved, which is much more significant than that of a TM mode waveguide. On the other hand, with sufficient variation of effective index but much lower loss, a TM mode waveguide is more suitable for construction of an electro-refractive phase modulator [24–26] .Note that these design principles for graphene-SiN$_x$ waveguides are quite different from graphene-Si waveguides, as quasi-TM modes exhibit stronger interaction over quasi-TE modes in Si-based structures [27]. This is due to the relative low refractive index contrast between SiN$_x$ and SiO$_2$, which results in a large substrate leakage loss [28]. In fact, a TE mode waveguide is more compatible with the on-chip optical integrated circuits because most laser diodes are operating with TE polarization [10].

The bending loss of an undoped graphene-SiN$_x$ waveguide as a function of radius of curvature is also obtained [29], as shown in Fig. 3(b). When the bending radius is larger than 35 μm, a limiting value in the order of 0.06 dB/μm is obtained, which is exactly the value of the loss of the straight graphene-SiN$_x$ waveguide we have solved above. Thus, we conclude that the bending loss of the microring structure is negligible for radius greater than 35 μm.

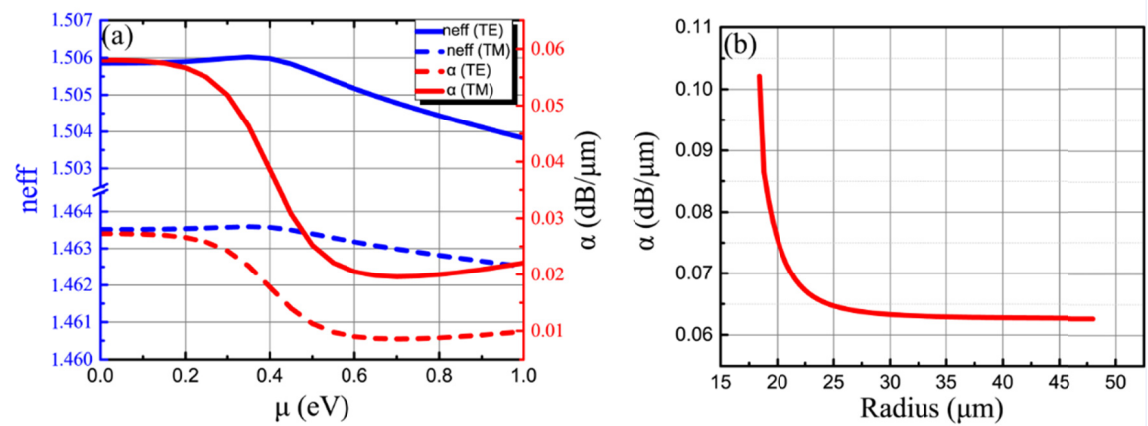

Fig. 3. (a) Effective index and propagation loss of the graphene-SiN$_x$ waveguide as a function of chemical potential of graphene. (b) Bending loss versus bending radius of the waveguide.

### 3. Optical modulators based on graphene-on-SiN$_x$ microring resonators
### 3.1 Theoretical analysis for microring resonator

To achieve a better performance, it's important to investigate the interaction between graphene and microring resonators. Here we study the influence of graphene induced electro-refractive phase modulation effect and electro-absorptive amplitude modulation effect [30], which both play an important role in the modulator. Considering an all-pass type microring resonator structure with graphene covering length of L, as schematically shown in Fig. 4. κ is the field cross coupling coefficient, r is the transmission coefficient of the coupling region and $\kappa^2 + r^2 = 1$ in a lossless coupler.

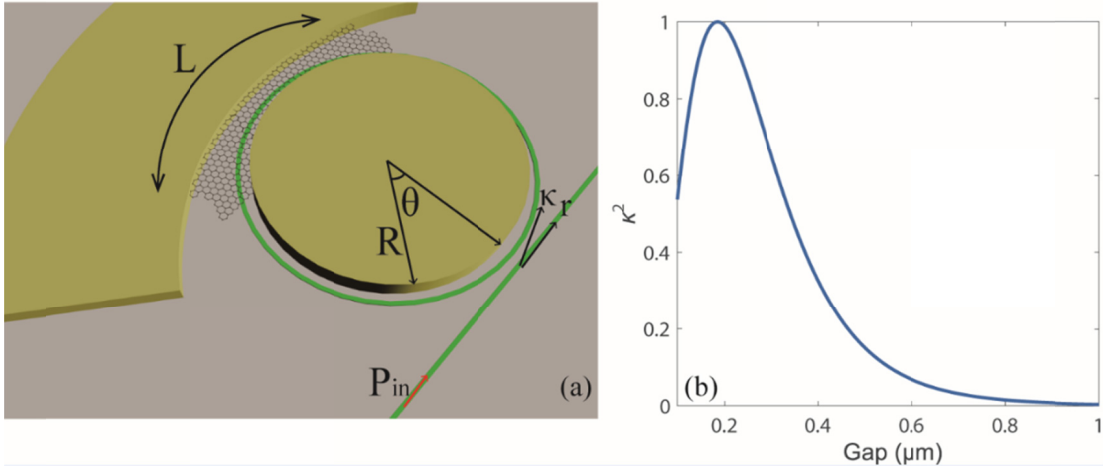

Fig.4 (a) Schematic illustration of the proposed graphene-on-SiN$_x$ microring resonator. (b) Dependency of κ on coupling gap of the microring resonator.

Coupling coefficient of a microring resonator can be derived based on the coupled mode theory (CMT) [31]. For a pair of waveguides with a width of 2a and a height of 2d, the mode coupling coefficient is described by:

$$K = \frac{\omega\varepsilon_0(n_1^2 - n_0^2)\int_{-a}^{a}\int_{-d}^{d}\vec{E}_{1x}^* \cdot \vec{E}_{1x}dxdy}{2\int_{-\infty}^{\infty}\int_{-\infty}^{\infty}\vec{E}_{1x}^* \cdot \vec{H}_{1y}dxdy} \quad (1)$$

$n_1$ and $n_0$ represent the refractive indices of waveguide and substrate, respectively. Thus the coupling coefficient for a microring resonator can be obtained by integrating over the entire coupling region,

$$\kappa = \sin(R \cdot \int_{-\frac{\pi}{2}}^{\frac{\pi}{2}} K \cdot \exp(-\gamma \cdot (g + 2R \cdot \sin^2\frac{\theta}{2})) \cdot \cos^2\theta \cdot d\theta) \quad (2)$$

where g is the coupling gap and $\gamma = \sqrt{\beta^2 - k_0^2 n_0^2}$. The coupling coefficient as a function of gap spacing is plotted in Fig. 4(b).

The loss coefficient of the graphene-SiN$_x$ microring can be written as:

$$a = \exp(-0.5\alpha_{GSN} \cdot L) \cdot \exp[-0.5\alpha_{SN} \cdot (2\pi R - L)] \quad (3)$$

where R is the radius of the microring, $\alpha_{SN}$ and $\alpha_{GSN}$ are the propagation loss of pure SiN$_x$ waveguides and graphene-on-SiN$_x$ waveguides, respectively. Here $\alpha_{SN}$ = 0.79 dB/cm is used [15]. Thus the transmission power is:

$$T = \frac{P_{out}}{P_{in}} = \frac{r^2 + a^2 - 2ar \cdot \cos\varphi}{1 + a^2 r^2 - 2ar \cdot \cos\varphi} \tag{4}$$

where $\varphi$ is the roundtrip phase shift and can be expressed as:

$$\varphi = \frac{\sum \omega_i L_i}{c} = \frac{2\pi \left[ n_{eff}^{GSN} \cdot L + n_{eff}^{SN} \cdot (2\pi R - L) \right]}{\lambda} \tag{5}$$

Then extinction ratio (*ER*) and insertion loss (*IL*) can be obtained by:

$$ER = 10 \log_{10} \frac{T_{max}}{T_{min}} = 10 \log_{10} \frac{(r+a)^2 (1-ra)^2}{(r-a)^2 (1+ra)^2} \tag{6}$$

$$IL = 10 \log_{10} T_{min} = 10 \log_{10} \frac{(r-a)^2}{(1-ra)^2} \tag{7}$$

**3.2 Numerical simulations and results**

According to the discussion above, an all-pass type microring resonator with a radius of 40 μm is studied. Resonant wavelength around 1.55μm determined by chemical potential of graphene and graphene coverage $(L/2\pi R)$ is estimated to 1.551±0.0005 μm. Thus we fix the wavelength at 1.551 μm and investigate the modulation depth (MD) at a high absorption state (*μ*=0.1 eV) and a low absorption state (*μ*=0.6 eV) with respect to graphene coverage and power cross coupling coefficient, which is shown in Fig. 5(a). Modulation depth is defined as $T(\mu = 0.6\,eV) - T(\mu = 0.1\,eV)$. For each coupling coefficient, there are two modulation depth peaks, which is significant when *κ²*>0.5. Thus a relatively high cross coupling coefficient *κ* can gives rise to high modulation efficiency.

Moreover, from the perspective of graphene-based device fabrication in practice, a high *κ* is preferable for modulator design. As a matter of fact, while fabrication of graphene-based devices seems simple, there are still a few inevitable problems, for instances, wrinkles, tears and pinholes on graphene usually exists [32] due to its single-atom-thickness. Thus it is necessary to consider the impact of fabrication tolerance on imperfect graphene coverage. According to theoretical analysis of microring resonator, the critical coupling condition is achieved by tuning the transmission coefficient *r* to balance the circulating loss coefficient *a*. Once chemical potential and coupling coefficient are settled, there is an optimum graphene coverage satisfying the critical coupling condition, which we defined as critical coverage. Critical coverage as a function of power coupling coefficient is plotted in Fig. 5(b). At a fixed coupling coefficient, graphene critical coverage monotonously decreases with decrease of chemical potential. For example, critical coverage for $\kappa^2$=0.5 decreases from 13% at µ=0.6 eV to 4.6% at µ=0.1 eV, suggesting the microring can always be tuned to critical coupling condition within this range by applying an appropriate voltage, thus an 8.4% fabrication tolerance of graphene coverage is acceptable. Obviously, a higher tolerance is obtained when circulating power in the microring increases. As a result, in practice, it is essential that graphene integrated microring resonators should be designed to have a relatively high coupling coefficient.

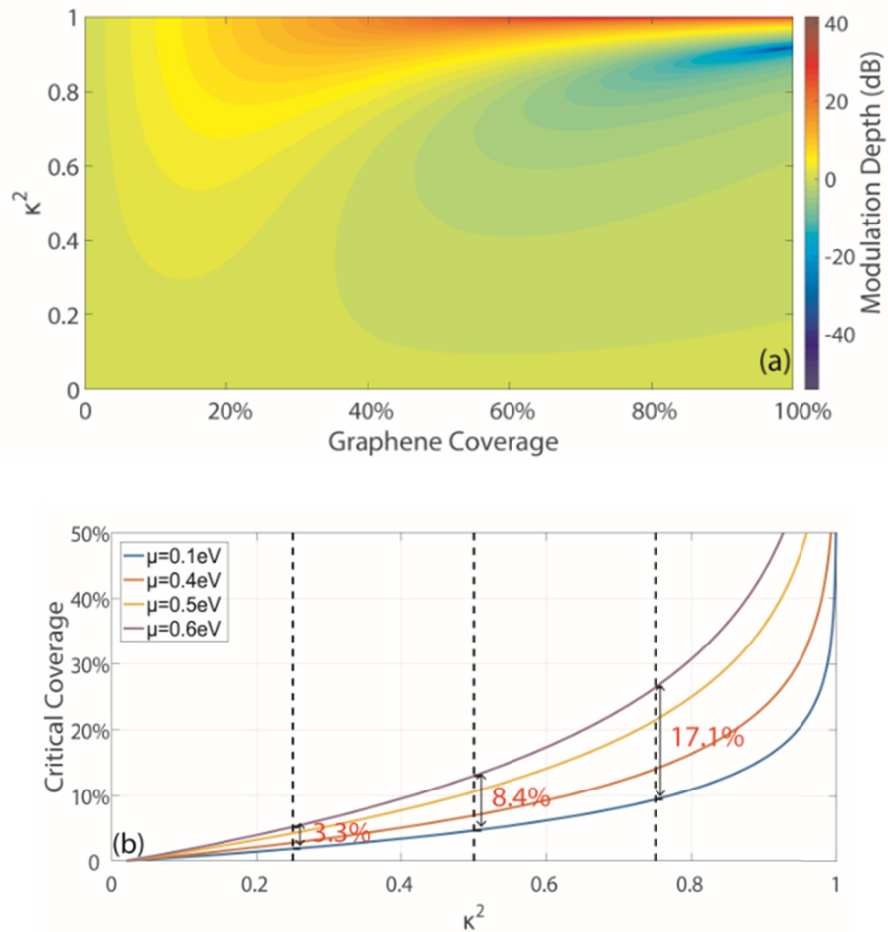

Fig. 5 (a) Modulation depth as a function of graphene coverage and power coupling coefficient. (b) Relationship between power coupling coeffficient and critical coverage.

Based on the analysis above, the coupling gap of the microring resonator is designed to be appropriately 120 nm, which corresponds to a high power coupling coefficient $\kappa^2$=0.75 with a 17.1% coverage tolerance. Then modulation depth and insertion loss with respect to graphene coverage and chemical potential (λ=1.551 µm) is investigated as shown in Fig. 6, which provides guidelines for choosing a suitable working point. An extremely high modulation depth can be obtained when graphene coverage is about 20% and chemical potential is tuned to about 1eV, which corresponds to a critical coupling point. However, one can see that from Fig. 6(b), insertion loss also reaches its maximum value when a high modulation depth is achieved. Thus a tradeoff between modulation depth and insertion loss should be considered.

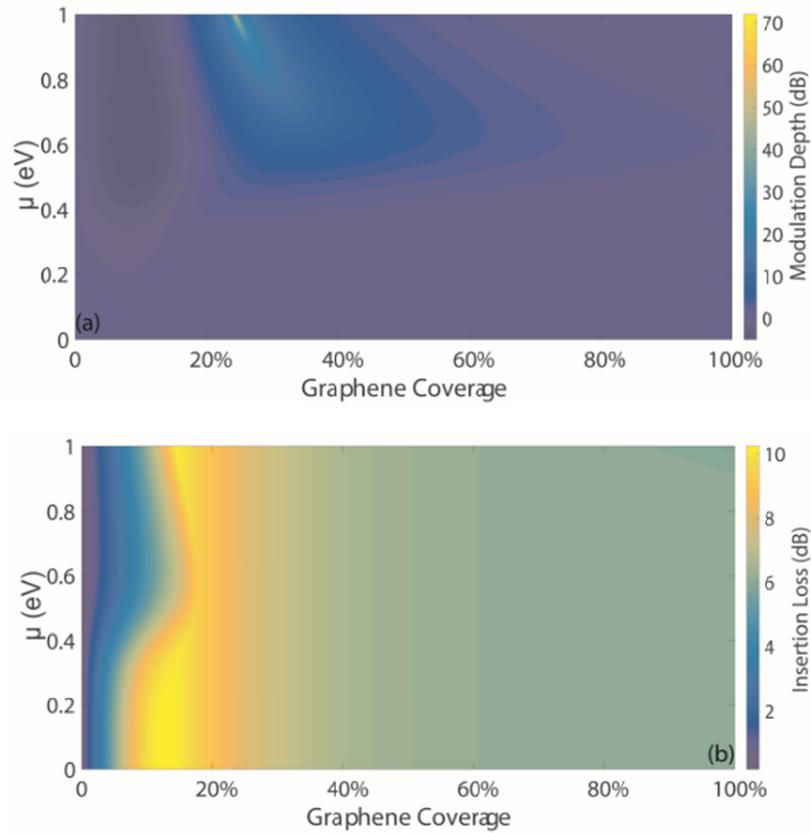

Fig. 6 (a) Modulation depth, and (b) insertion loss as a function of chemical potential of graphene and graphene coverage

Finally, we simulated the transmission spectra of different chemical potentials at fixed coverage, as exhibited in Fig. 7. Note that the values of coverage 14%, 21.4% and 26.3% are calculated by Eq. (4), which are critical coverage for $\mu$ = 0.4 eV, $\mu$ = 0.5 eV, $\mu$ = 0.6 eV. The insets in Fig. 7(b), (d) and (f) show the resonance wavelength and corresponding modulation depth(MD) and insertion loss(IL). On the basis of the theoretical analysis, the microring resonator can always be tuned into resonance from 14% coverage to 26.3% coverage, indicating that we can select a 20% coverage with a large fabrication tolerance of 12% (i.e., ±6%) for design. When the resonance is set at a high absorption working point ($\mu$ = 0.4 eV), insertion loss is rather significant. By increasing graphene coating length the resonance point can be modified to a low absorption state, achieving a lower insertion loss. Besides, larger graphene coverage results in a lower quality factor, so photon lifetime will not affect device bandwidth [18].

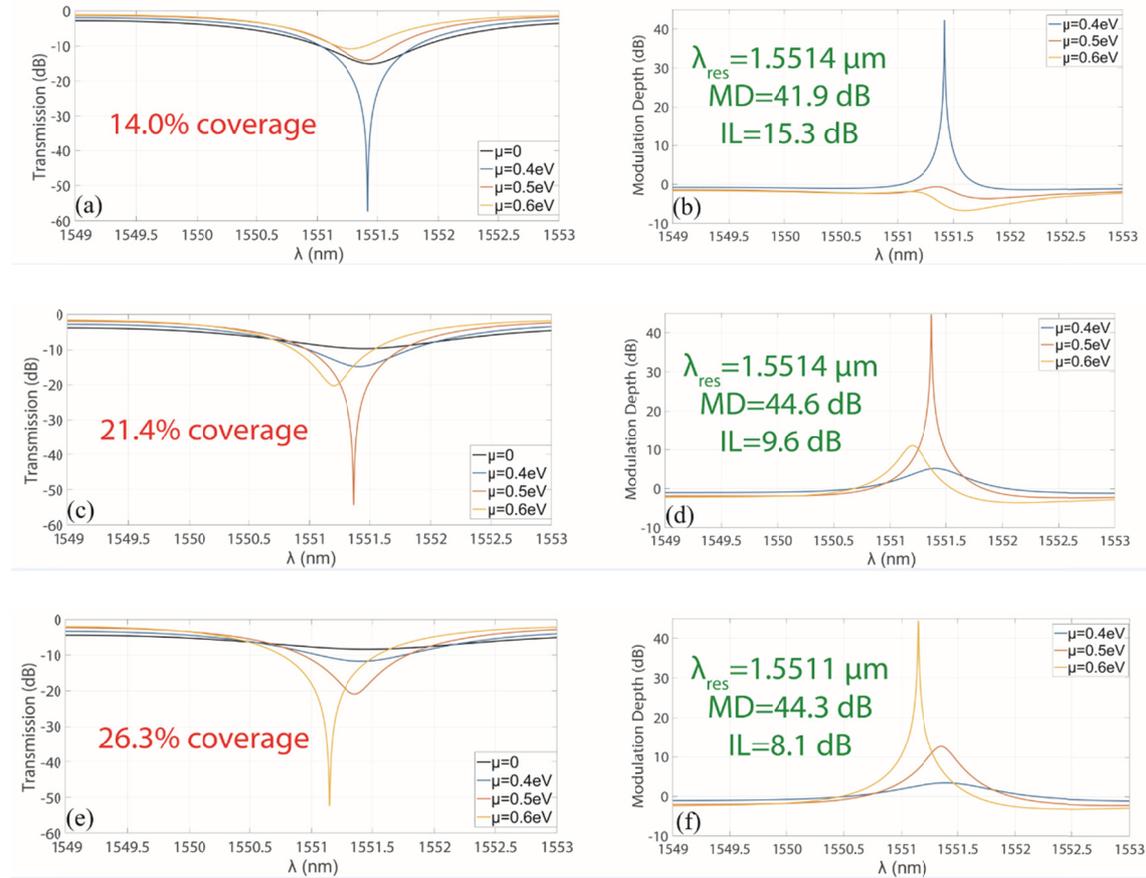

Fig. 7 Simulated transmission spectra and corresponding modulation depth with (a) (b) 14.0% , (c) (d) 21.4% and (e) (f) 26.3% coverage of graphene for different chemical potential.

## 4. Conclusion

In summary, we investigate the tunable loss and effective index in graphene-on-SiNx waveguides which exploits graphene's controllable optical conductivity. It is found that TE polarization mode is more suitable for graphene-SiN$_x$ modulators, with which a loss tunability of 0.04 dB/μm and an effective index variation of 0.0022 can be achieved after optimizing waveguide dimensions. Using microring bending radii of larger than 35 μm, where bending loss is insignificant compared to the absorption loss of graphene, graphene-on-SiN$_x$ microring modulators are comprehensively optimized considering the impacts of coupling coefficient and chemical potential on modulation depth and insertion loss, providing important instructions for the design of graphene-based microring resonator. Simulation results indicate that with a relatively high power coupling coefficient κ²=0.75 and proper graphene coverage, multiple points of operation can be found. Exemplary designs with high modulation depth of ~40 dB can be obtained from a microring modulator of 40 μm radius with a graphene coverage ranging from 14% to 26.3%.

**Acknowledgment**

This work is supported by National Basic Research Program of China (973 Program) (2014CB340000, 2012CB315702); Natural Science Foundations of China (61323001, 61490715, 51403244, 11304401); Natural Science Foundation of Guangdong Province (2014A030313104); Fundamental Research Funds for the Central Universities of China (Sun Yat-sen University: 13lpgy65, 15lgpy04, 15lgzs095, 15lgjc25, 16lgjc16); Specialized Research Fund for the Doctoral Program of Higher Education of China (20130171120012).